# On the qualitative difference between phonon-assisted transition rates of electrons and holes in organic disordered semiconductors.

E.D. Gutliansky[1*], Nir Tesler[1] and Uri Peskin[2]

[1]Zisapel Nano-Electronic Center, Department of Electrical Engineering, Technion - Israel Institute of Technology, Haifa 32000, Israel.

[2]Schulich faculty of Chemistry, Technion – Israel institute of Technology, Haifa, 32000, Israel

We suggest universal expressions for the rates of charge transition between molecules in organic disordered semiconductors, which differ between electrons and holes. The donor and acceptor molecules (monomers) are represented in terms of their frontier orbitals, with asymptotic tails reflecting the different asymptotic potential wells which bind different charge carriers. This model predicts an asymmetry of the tunnel factors and the interaction with phonon field with respect to electrons and holes, leading to different formulas for the transition rates and the thermal mobility of the different charge carriers. This is demonstrated explicitly for transitions between organic molecules initiated by fluctuations of the inter-molecular distance due to deformation of by the phonon field, but the asymmetry between the carriers is universal and should be essential for quantitative analysis of charge transport in any disordered material. In particular, in the framework of classical Mott considerations our result predict an unusual temperature dependence of holes mobility, $\sim (T_0/T)^r \exp-(T_0/T)^{1/4}$, where $r = \left(E_F^h\right)^{-1/2}$, and $E_F^h$ is the energy of hole at the Fermi level in atomic units.

Organic disordered semiconductors have been developed for a range of electronic devices such as light-emitting diodes [1]-[2], field effect transistors [3], photoreceptors [4] and photovoltaic cells [5]. An understanding of the carrier mobility in an organic semiconductor is necessary in order to optimize the performance of polymeric electronic devices. The macroscopic mobility is controlled by the density of states, the morphology, and mainly by the charge transition rate between donor and acceptor molecular pairs.

The latter is dominated to a large extent by the frontier molecular orbitals: highest occupied molecular orbital (HOMO) or lowest unoccupied molecular orbital (LUMO) [6]. Much discussion was devoted to fluctuations in the molecular orbital energies in the disordered crystal environment. These are typically one order of magnitude larger than the corresponding charge transfer integrals[10], implying that the carrier wave functions can be considered strongly localized [7] – [10] and molecules or molecular subunits can be considered as point sites [7]–[11]. The problem of calculating the transition rate between two sites therefore becomes similar to the well studied problem of transitions between donors in doped crystalline semiconductors, which was solved by Miller and Abraham (MA) [12]. Although applicable to charge transport in disordered organic and inorganic semiconductors [13] – [26], this approach was criticized on the basis of comparison between MA calculations and the results of detailed atomistic level calculations [27]. Indeed, different aspects related to the application of MA theory to organic materials need to be revisited. First, in MA theory the transition rates for electrons and holes have the same mathematical form, and differ from each other only by the "delocalization lengths" which is a phenomenological parameter of the theory, independent on the donor and acceptor orbital energies. However, modern experiments show that the mobility of electrons and holes may be similar or different by orders of magnitudes in organic [28]-[29] and in inorganic materials [30]. Second, the MA rate expression was deduced where the interaction of the charge with the phonon field was due to zone deformation. This is well suited for an electron donor and an acceptor which are parts of a classical semiconductor lattice, but in disordered organic materials this is not the case, and other forms of interactions between the charged molecules and the phonons should be considered as triggers for charge hopping events. Indeed, typical organic electronic devices are made of thin organic films deposited on a glass substrate, where the elastic moduli of the substrates is much larger than the elastic moduli of the organic films. The thickness of films is of the order of 100 nm, which is much smaller than the thickness of substrate. Therefore, oscillations in the film will be controlled by the surface acoustic waves (SAW) of the substrate [31]. We consider these oscillations to be the main perturbation which modulates the distances between molecules and facilitates transitions of electrons between them. Notice that these acoustic phonons are associated with wave lengths much larger than the typical molecular size. Electron phonon interactions associated with localized internal molecular vibrations (short wave-length phonons which engender polarons of small radius [32]) are therefore excluded in our treatment below. This phonon-induced transport mechanism is therefore different from the zone deformation considered in the classical works of Miller Abraham [12], and from the on-site energy fluctuations considered by Holstein [32] and Emin[33].

The goal of the present work is the derivation of new formulas for the transition rates of electrons and holes in disordered materials. As a concrete example, we consider the modulation of the inter-molecular distance by deformation of the phonon field [31] which should be a prominent mechanism for charge transport in disordered organic materials.

The major new outcomes which go beyond the MA formula [12] are as follows:1. Our result gives simple dependence of the tunneling matrix element on the electronic orbital energies of the donor and the acceptor.

2. The transition rate for electrons and holes differ not only parametrically, due to localization length, but also in their functional form, owing to different forms of the asymptotic electronic potential for an electron and a hole. This result explains the observation that the mobility of holes and electrons can differ only slightly, although their localization length greatly differ, a fact that can't be explained by MA theory.

When considering charge transfer between molecules at thermal conditions, it is usually sufficient to consider the frontier molecular orbital at each site, characterized by its orbital energy and localization length. A model which accounts for these properties is a set of effective potential wells. Each molecule (monomer unit) is modeled as a single potential well (PW) and polymers or clusters of molecules (monomers) are considered as aggregates of potential wells. Below we assume that each PW supports an isolated energy level at the energy of the relevant molecular orbital taking part in the electron transfer process (HOMO or LOMO). Since the distances between PWs typically exceed by an order of magnitude the orbitals localisation radius in the PWs, the transition rates are insensitive to the details of the PW shape, and are controlled primarily by their asymptotic form. Indeed, the asymptotic potential confining the charge in the respective potential well determines the asymptotic tail of the molecular orbital, and thus, the overlap between orbitals of different molecules.

The effective PWs for electron transfer and for hole transfer have quite different asymptotic forms. In the case of electron hopping between two LUMOs, associated with two PWs, the occupied PW is negatively charged, whereas the unoccupied PW is electro neutral (See Fig. 1a). The asymptote of each PW corresponds therefore to the attraction of an electron to an electro-neutral molecule, dominated by the charge-dipole interaction, which scales like $-1/r^4$. In the case of electron hopping between two HOMOs (or hole hopping), the occupied PW is electro neutral, while the unoccupied PW is positively charged (See Fig. 1b). The asymptote of each PW should therefore have the form of an attractive Coulomb ($-1/r$) potential in this case. The effective model Hamiltonian describing charge hopping between two potential wells therefore reads,

$$\hat{H}^{h,e} = -\frac{1}{2}\Delta + V_a^{h,e}(\vec{r}-\vec{r}_a) + V_b^{h,e}(\vec{r}-\vec{r}_b),\tag{1}$$

where the electron ($e$) or hole ($h$) potentials have different asymptotic forms, at $r \to \infty$,

$$V_{a,b}^{h}(\vec{r}-\vec{r}_{a,b}) \to -\frac{1}{|\vec{r}-\vec{r}_{a,b}|} \qquad V_{a,b}^{e}(\vec{r}-\vec{r}_{a,b}) \to -const/|\vec{r}-\vec{r}_{a,b}|^4.\tag{2}$$

Each PW is assumed to support an isolated bound state, at the respective orbital energies, $E_a^{h,e}, E_b^{h,e}$, where the energy difference between the two potential wells is typically larger than the respective tunneling matrix element. Without loss of generality, we shall assume here, $E_b^{h,e} > E_a^{h,e}$, such that a transition of a charge carrier from site a to site b is in fact a transition from the ground state into the first exited state of the Hamiltonian (Eq.1)

The corresponding ground ($\Psi_{gr}^{h,e}$) and excited ($\Psi_{ex}^{h,e}$) wave functions of Eq. (1) can be approximated as follows,

$$\Psi_{gr}^{h,e} \cong \frac{1}{\sqrt{1+k^2}}\left[\psi_a^{h,e}(\vec{r}-\vec{r}_a) + k\psi_b^{h,e}(\vec{r}-\vec{r}_b)\right]$$

$$\Psi_{ex}^{h,e} \cong \frac{1}{\sqrt{1+k^2}}\left[-k\psi_a^{h,e}(\vec{r}-\vec{r}_a) + \psi_b^{h,e}(\vec{r}-\vec{r}_b)\right],\tag{3}$$

Here, $\psi_a^{h,e}, \psi_b^{h,e}$ are the localized orbitals associated with electron localization on the molecular sites, a or b, respectively, where the overlap between them is assumed to be negligible ($\int \psi_a^{h,e}\psi_b^{h,e}d\vec{r} \approx 0$) in view of the large ratio between the inter-molecular distances and the orbitals localization length. The parameter k can be expressed explicitly following Ref.[34]. Starting from an exact expression,

$$2\left(E_I^{h,e} - E_{II}^{h,e}\right)\int_V \Psi_I^{h,e}\Psi_{II}^{h,e}d\vec{r} = \oint_S (\Psi_I^{h,e}\nabla\Psi_{II}^{h,e} - \Psi_{II}^{h,e}\nabla\Psi_I^{h,e})d\vec{s}, \qquad (4)$$

where $\Psi_I, \Psi_{II}$ and $E_I, E_{II}$ are exact eigenfunctions and eigenvalues of the Hamiltonian $\hat{H}^{h,e}$, we substitute $\Psi_{ex}^{h,e}, \Psi_{gr}^{h,e}$ instead of $\Psi_I^{h,e}, \Psi_{II}^{h,e}$, and $-E_{ex}^{h,e}, -E_{gr}^{h,e}$ instead of $E_I^{h,e}, E_{II}^{h,e}$. Integrating over the volume associated with site 'a', i.e., $V_a$, and neglecting overlap integrals and the corresponding penetration integrals, $\int_{V_{a,b}} |\psi_{b,a}^{h,e}|^2 d\vec{r} \approx 0$, one obtains,

$$\frac{k^{h,e}}{1+(k^{h,e})^2} = \frac{\oint_S \left(\psi_a^{h,e}\nabla\psi_b^{h,e} - \psi_b^{h,e}\nabla\psi_a^{h,e}\right)d\vec{s}}{2(E_b^{h,e} - E_a^{h,e})}. \qquad (5)$$

For calculation of the numerator of the right hand side of Eq.(5) it is comfortable to choose a plane S that is perpendicular to the line connecting the two sites, which divides this line in half. Using cylindrical system of coordinates we can find,

$$\frac{k^{h,e}}{1+(k^{h,e})^2} = \frac{\pi R \psi_a^{h,e}(R/2)\psi_b^{h,e}(R/2)}{\left(E_b^{h,e} - E_a^{h,e}\right)} \quad ; \quad R = \left|\vec{r}_a - \vec{r}_b\right| \qquad (6)$$

The local site functions, $\psi_a^{h,e}$ and $\psi_b^{h,e}$, are different for the electron and the hole potential wells. In the case of electron transfer, the asymptotic wave function for the a,b sites is proportional to the eigen function of the respective effective single particle potential, and reads[35],

$$\psi_{a,b}^e(\vec{r} - \vec{r}_{a,b}) = \frac{A_e}{\left|(\vec{r} - \vec{r}_{a,b})\right|}\exp(-\alpha_{a,b}^e\left|(\vec{r} - \vec{r}_{a,b})\right|) \qquad (7)$$

where, $\dfrac{(\alpha_{a,b}^{h,e})^2}{2} = E_{a,b}^{h,e}$. For hole transfer, the asymptotic eigen function of the effective single particle local potential well has the form [34-37]

$$\varphi_{a,b}^h\left(\left|\vec{r}-\vec{r}_{a,b}\right|\right) = A_h \left|\vec{r}-\vec{r}_{a,b}\right|^{\frac{1}{\alpha_{a,b}^h}-1}\exp\left(-\alpha_{a,b}^h\left|\vec{r}-\vec{r}_{a,b}\right|\right). \qquad (8)$$

Introducing a correction to the local function due to the deformation induced by the electric field at the neighboring site (in the $z$ direction) [38], the asymptotic hole function reads,

$$\psi_{a,b}^{h}(\vec{r} - \vec{r}_{a,b}) = \left(\frac{R}{R - |z|}\right)^{\frac{1}{\alpha_{a,b}^{h}}} A_h \left|\vec{r} - \vec{r}_{a,b}\right|^{\frac{1}{\alpha_{a,b}^{h}} - 1} \exp\left(-\alpha_{a,b}^{h}\left|\vec{r} - \vec{r}_{a,b}\right|\right). \tag{9}$$

$A_h$ and $A_e$ are dimensionless parameters of the order of unity which depend on the internal structure of the potential wells (internal electron shells [34]-[37], and therefore depend weakly on the energy. Substituting Eqs. (7,9) in Eq.(6) we obtain,

$$\frac{k^e}{1 + (k^e)^2} = 4\pi A_e^2 \left(E_b^e - E_a^e\right)^{-1} R^{-1} \exp[-(\alpha_a^e + \alpha_b^e)\frac{R}{2}] \tag{10}$$

$$\frac{k^h}{1 + (k^h)^2} = 4\pi A_h^2 \left(E_b^h - E_a^h\right)^{-1} R^{\frac{1}{\alpha_a^h} + \frac{1}{\alpha_b^h} - 1} \exp[-(\alpha_a^h + \alpha_b^h)\frac{R}{2}] \tag{11}$$

The last results enable to express the ground and first excited eigenfunctions of the two PWs Hamiltonians in terms of the respective energies and the inter-molecular distance.

Let us turn to the formulation of the interaction which induces transition between these states. The perturbation Hamiltonian which induces electron transitions between the two states takes the following form:

$$\hat{H}_{int} = \frac{\partial}{\partial \vec{R}}\left[V_a(\vec{r}) + V_b(\vec{r} - \vec{R})\right]\bigg|_{\vec{R} = \vec{R}_0} \Delta \vec{R}. \tag{12}$$

Here, $\vec{R}_0 = \vec{r}_b - \vec{r}_a$ is the distance vector between the two potential well minima at the equilibrium configuration (the origin of the coordinate system was set to $\vec{r}_a$), and $\Delta \vec{R} \equiv [(\vec{R} - \vec{R}_0) \cdot \frac{\vec{R}_0}{R_0}]\frac{\vec{R}_0}{R_0}$ is the deviation from the equilibrium distance, in the direction of $\vec{R}_0$. $\Delta \vec{R}$ can be expressed in terms of the deformation vector ($U$) of the surface substrate. Since the main contribution to this deformation is attributed to surface acoustic waves (SAW), whose wave length is much longer than the inter-molecular distance ($\lambda \gg |\vec{R}_0|$), the value of $U$ is approximately constant at any point on the distance vector between the two PWs. For definiteness we set

$$\Delta \vec{R} = \left.\frac{\partial U}{\partial \vec{R}}\right|_{\vec{R}_0} \vec{R}_0 , \qquad (13)$$

where $U = (\vec{U} \cdot \vec{R}_0)/R_0$ is the longitudinal component of the AW in the direction of $\vec{R}_0$, and $R_0 = |\vec{R}_0|$. This component of the deformation vector in the second quantization representation has a form (see for example [39],[12]):

$$U(\vec{R}) = \sum_q \sqrt{\frac{\hbar}{2V\rho_0 C q}} \cos\theta \left[ b_q^- \exp(i\vec{q}\vec{R}) + c.c. \right]. \qquad (14)$$

Here $\rho_0, V, C$ are respectively the density of substrate states, its volume, and the velocity of the AW; $b_q^+$ and $b_q^-$ are the creation and annihilation operators for AW phonons of wave vector $\vec{q}$, $q = |\vec{q}|$, and $\theta$ is the a angle between $\vec{R}_0$ and $\vec{q}$. Substituting Eqs.(13,14) in Eq.(12), we obtain,

$$\hat{H}_{int} = iR_0 \left.\frac{\partial}{\partial \vec{R}}\left[V_b(\vec{r}-\vec{R})\right]\right|_{\vec{R}=\vec{R}_0} \sum_q \sqrt{\frac{\hbar q}{2V\rho_0 C}} \cos\theta \left[ b_q^- \exp(iqR_0 \cos\theta) + c.c. \right]. \qquad (15)$$

Without loss of generality, let us consider a transition from PW 'a' to PW 'b', assuming $E_a^{h,e} \prec E_b^{h,e}$. Such transitions are induced by phonon absorption due to the (weak) interaction between the electron and the AW phonon field. Using the above formulation, the transition amplitude is nearly equal to the amplitude of transition between the respective eigenstates of Eq.(1), i.e.,

$$H_{a\to b}^{h,e} = \left\langle \Psi_{gr}^{h,e} \left| \hat{H}_{int} \right| \Psi_{ex}^{h,e} \right\rangle$$

$$= iR_0 \sqrt{\frac{\hbar q n_q}{2V\rho_0 C}} \frac{k^{h,e}}{1+(k^{h,e})^2} \cos\theta \exp(iqR\cos\theta) \qquad , \qquad (16)$$

$$\left( \left\langle \psi_b^{h,e} \left| \left.\frac{\partial}{\partial \vec{R}}\left[V_b(\vec{r}-\vec{R})\right]\right|_{\vec{R}=\vec{R}_0} \right| \psi_b^{h,e} \right\rangle - \left\langle \psi_a^{h,e} \left| \left.\frac{\partial}{\partial \vec{R}}\left[V_b(\vec{r}-\vec{R})\right]\right|_{\vec{R}=\vec{R}_0} \right| \psi_a^{h,e} \right\rangle \right)$$

where we have neglected terms involving multiplications between the localized wave functions $\psi_a^{h,e}$ and $\psi_b^{h,e}$, in view of their negligible overlap (considering that $\alpha_a^{h,e} R_0, \alpha_b^{h,e} R_0 \gg 1$). $n_q$ is the thermal phonon occupation number in the mode which matches the transition energy, $n_q = 1/[\exp(|E_a^{h,e} - E_b^{h,e}|/(K_B T)) - 1]$. Notice that for the case of the inverse transition from 'b' to 'a' (corresponding to phonon emission), one needs to change: $n_q \to n_q + 1$.

The first term in Eq.(16) vanishes for both electron and hole transitions, because both the wave function $\psi_b^{h,e}$, and the PW $V_b^{h,e}$ are centered around the point $\vec{r} - \vec{R} = 0$. The second term in Eq.(16) is different for holes and electrons because of the difference in the asymptotes of the potential wells. For hole transfer,

$$\left\langle \psi_a^h \left| \frac{\partial}{\partial \vec{R}} \left[ V_b^h (\vec{r} - \vec{R}) \right] \right|_{\vec{R}=\vec{R}_0} \left| \psi_a^h \right\rangle \approx -\frac{e^2}{R_0^2}, \tag{17}$$

since the asymptotic potential has a Coulomb form, whereas for electron transfer, the asymptotic interaction with the electroneutral atom is dominated by the induced dipole moment, $p \sim \alpha e/|\vec{r} - \vec{R}|^2$, ($\alpha$ is the monomer or small molecule polarizability), and the asymptotic potential for the electron takes the form, $V_b^e(\vec{r} - \vec{R}) \sim -2\alpha e^2/|\vec{r} - \vec{R}|^4$. Consequently, in analogy to the case of an atom [35], one obtains,

$$\left\langle \psi_a^e \left| \frac{\partial}{\partial \vec{R}} \left[ V_b^e (\vec{r} - \vec{R}) \right] \right|_{\vec{R}=\vec{R}_0} \left| \psi_a^e \right\rangle \approx 8\frac{e^2 \alpha}{R_0^5} \tag{18}$$

Using Eqs.(17,18) and defining, $\gamma_h = \frac{e^2}{R_0}$, and $\gamma_e = \frac{\alpha e^2}{R_0^4}$, the transition amplitude (Eq. 16) can be written as follows,

$$H_{a \to b}^{h,e} = -i\gamma_{h,e} \sqrt{\frac{\hbar q n_q}{2 V \rho_0 C}} \frac{k^{h,e}}{1+(k^{h,e})^2} \cos\theta \exp(iqR\cos\theta) \tag{19}$$

The parameters $\gamma_h$ and $\gamma_e$, play the role of the electron phonon-interaction strength parameters, as in the theory of MA [12]. Notice that our treatment predicts that the interaction coefficient is always much larger for holes than for electrons since $\gamma_e/\gamma_h = \alpha/R_0^3$. Considering that the molecular polarizability $\alpha$ scaled like the cube of the effective molecular radius, which is much smaller than the distance between the PWs, one has, $\gamma_e/\gamma_h \ll 1$.

The phonon-assisted transition rate of an electron or a hole from PW 'a' to PW 'b', are expressed in terms of the transition amplitude using Fermi's Golden rule

$$W_{a \to b}^{h,e} = (2\pi/\hbar)(V/8\pi^3) \int \left| H_{a \to b}^{h,e} \right|^2 \delta\left[\hbar C q - \Delta_{h,e}\right] d\vec{q} \quad , \tag{20}$$

where, $\Delta_{h,e} = (me^4/\hbar^2)(E_b^{h,e} - E_a^{h,e})$, (not restricted to atomic units) and using Eq.(20), we get the transition rates for holes and electrons:

$$W_{a \to b}^h \propto A_h^2 \frac{\gamma_h^2}{\rho_0 C^5 \hbar^4} \left| \Delta_h \right| (me^4/\hbar^2)^2 n_q R^{\frac{2}{\alpha_a^h} + \frac{2}{\alpha_b^h} - 2} \exp[-(\alpha_a^h + \alpha_b^h)R] \tag{21}$$

$$W_{a \to b}^e \propto A_e^2 \frac{\gamma_e^2}{\rho_0 C^5 \hbar^4} \left| \Delta_e \right| (me^4/\hbar^2)^2 n_q R^{-2} \exp-[(\alpha_a^e + \alpha_b^e)R] \tag{22}$$

The last two equations are central results of the present treatment, as they show that the difference between the rates of electron and hole transfer are not restricted to the parameters values ($\gamma$, $A$). The functional dependence on the inter-site distance is different, where hole transfer rate seems to drop slower with distance. Moreover, the precise distance dependence is shown to depend on the electron binding energy at the respective frontier orbitals (through $\alpha_a, \alpha_b$) suggesting a non-trivial relation between the transport rates of electrons and holes in a given material.

Let us discuss now the regime of applicability of the above formulas for interpretations of conductivity in disordered organic semiconductors. In the case of thin films, the origin of phonon assisted charge transport is the deformation of the substrate surface. This is a collective motion of many substrate atoms, which is attributed to phonon degrees of freedom. Organic films are, in most cases, deposited on a ceramic (glass) substrate, with a Debye temperature, e,g,, 495K [40].

Since this value exceeds room temperature (~300K), the surfaces phonons [41] must be treated as a quantum field at room temperature, and the above consideration for the transition rates can be applied. It is important to note in this context that in some work for calculating mobility, Marcus expression is used, see for example [28],[42]-[43]. In these cases the necessary energy needed to compensate the difference in energies between the electron on the donor and on the acceptor molecules is taken from local internal fluctuations of degree of freedom of the donor molecule. This is very different from the mechanism considered here, where the source of energy is the long-range deformation field. Notice that we do not consider here the case of flexible substrates, where the film would be deposited on plastic. Nevertheless, our approach is easy generalized for that case as well.

Finally, let us consider the simplest implementation of the above rate formulas for calculations of the macroscopic mobilities of the different charge carriers. For electron transport we will consider the Fermi level to lay in the LUMOs zone, and for hole transport, in the HOMOs zone. In each case we shall invoke the Mott assumptions, i.e. a constant density of states at Fermi level, and $|E_b - E_a|/E_{b,a} \ll 1$, at room temperature. Using the variable rang hopping approximation [13] for the mobility calculation we have to find the maximal transition rate $W_{opt}^{h,e}$ (21),(22) and the distances, $R_{opt}$, at which they are obtained. Analysis of the corresponding trancendental equations shows that in this approximation the dispersion of the localization radius has a small effect on the temperature-dependence of the mobility. Therefore, in our illustrative consideration we will not take into account the energy dependence of the localization radius. Using the approximation $\left[ 2^{1/2} kT / \left( 2 E_F^{h,e} \right)^{3/2} \right] \ln R \ll 1$, (here $E_F^{h,e}$ is the Fermi level tuned to the HOMO or LUMO orbitals), and using the expression for the mobility $\mu = e W_{opt}^{h,e} R_{opt}^2 / 6kT$, we find that the temperature dependence of the mobility for holes and electrons reads, $\mu^h \sim \left( T_0^h / T \right)^{r^h} \exp - \left( T_0^h / T \right)^{1/4}$, and $\mu^e \sim \exp - \left( T_0^e / T \right)^{1/4}$, respectively. Here $T_0^h = (512/9\pi)(r^h a_0)^{-3} (N_F k)^{-1}$, $r^{h,e} = \left( E_F^{h,e} \right)^{-1/2}$, and, $a_0 = 5.29 \cdot 10^{-1}$ Å. While the Mott approximation for the temperature-dependence of the mobility for electrons is reproduced, the hole mobility accumulates an extra power term. In order to understand the possible role of this power term in the hole mobility let us evaluate it approximately. Considering the overlap parameter from

Ref. [44], $r^h a_0 = 1.6$ Å, hence, $r^h = 3.6$, and setting[45], $N_F = 3.2 \cdot 10^{29} m^{-3} ev$, we obtain for $T_0^h \approx 10^8 K$, and for the mobility of holes, $\mu^h \sim \left(T_0^h/T\right)^{3.6} \exp-\left(T_0^h/T\right)^{1/4}$. This effect on the mobility should be observable experimentally. For a quantitative prediction, however, the inhomogeneous density of states and the specific distribution of donor-acceptor energy mismatches must be accounted for.

In conclusion, we suggest a universal model for charge transport in disordered materials which takes into account the asymptotic peculiarity of the frontier orbitals (HOMO and LOMO) of the charge donor and acceptor, where the source of energy for the transition is fluctuations in the donor-acceptor distance induced by the phonon field. This model gives new simple formulas for transition rates which qualitatively differ between electron and holes and take into account the orbital energies of the donor and acceptor. We have shown, in particular, that it leads to a qualitative change of the classical Mott result for disordered materials. This result indicates the need to revise the existing theory of transport in disordered materials. Although the model is suggested for transition between two molecules (monomers) it can be easily generalized for description of the transition of charge carriers between polymers where several monomers or molecules participate in the transport simultaneously. This extension would give a possibility to calculate the mobility also in polymeric materials, taking into account their morphology. We emphasize that the difference between electron and hole mobilities as demonstrated here should be applicable for disordered materials in general, where the transport is controlled by the hopping mechanism. The model substantially advances the understanding of transport processes in disordered materials by accounting for the fundamental difference in mobility between electrons and holes, and opens a significant new area of theoretical and experimental research aiming to characterise the difference in the mobility of electrons and holes in various disordered materials, on the base of our model.

ACKNOWLEDGMENTS

Work of one from us (E. Gutliansky) partly was supported by, The Center for Absorbtion in Science, Ministry of Immigrant Absorption, State of Israel.

*Electronic address: e.d.gutliansky@gmail.com

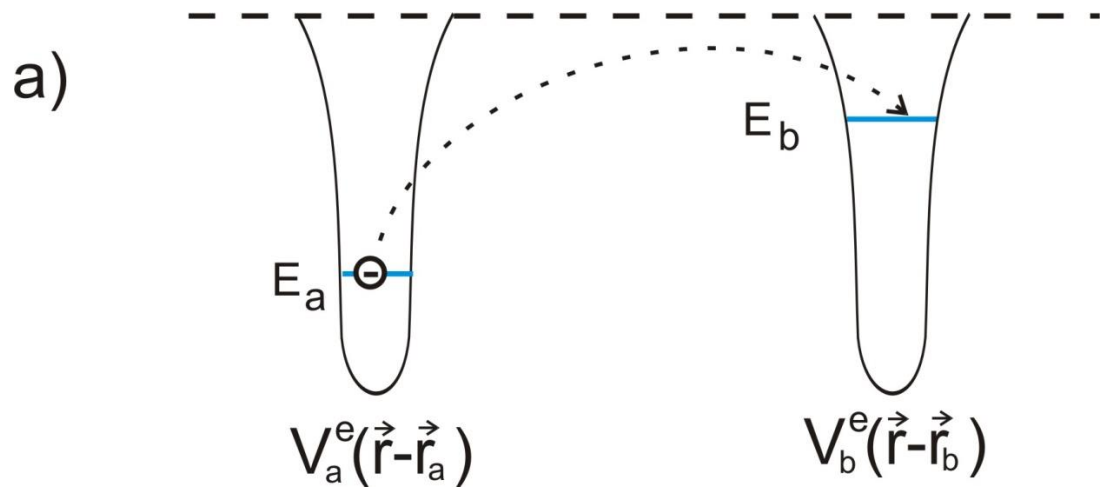

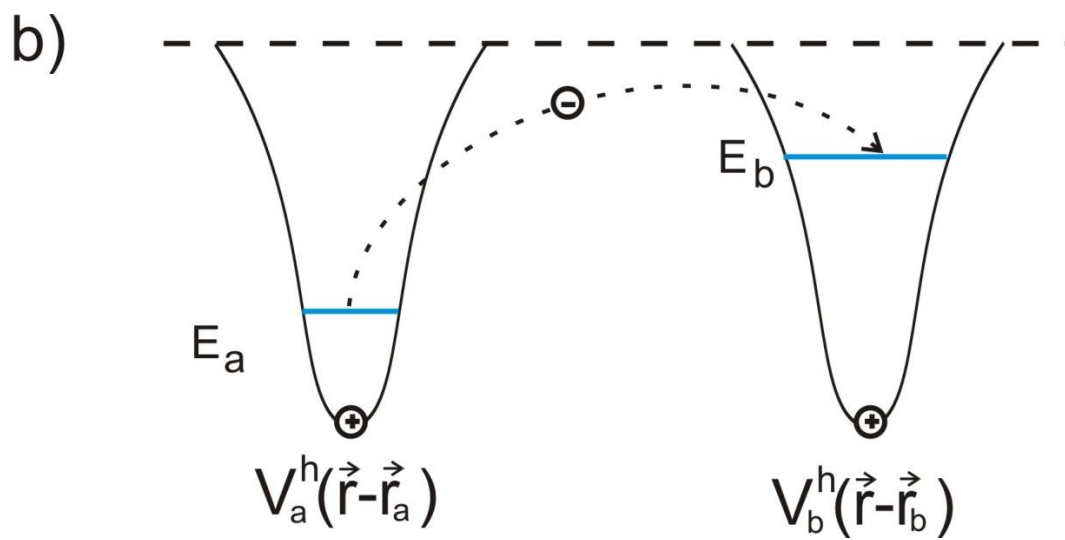

Fig.1. a) Transition of an electron between electro neutral potential wells, with charge-dipole asymptotic interaction. This model corresponds to electron hopping between lowest unoccupied molecular orbitals. b) Transition of an electron between positively charged potential wells, with Coulomb asymptotic interaction. This model corresponds to hole hopping between highest occupied molecular orbitals.